\documentclass[referee,floatfix]{aa} % for a referee version
\usepackage{graphicx}
\begin{document}
%\textbf (\bfseries)
   \title{On Shadow of the Moon in Extensive Air Shower Data}

%  \subtitle{}

   \author {F. Sheidaei
          \inst{b} \fnmsep \thanks{\email{sheydaei@sharif.ir}}
          \and
           A. Anvari
          \inst{a,b} \fnmsep \thanks{\email{anvari@sharif.ir}}
          \and
           M. Bahmanabadi
          \inst{a,b} \fnmsep \thanks{\email{bahmanabadi@sharif.ir}}
          \and
        M. Khakian Ghomi
          \inst{a} \fnmsep \thanks{\email{khakian@sharif.ir}}
          J. Samimi
          \inst{a,b} \fnmsep \thanks{\email{samimi@sharif.ir}}
         }

   \offprints{F. Sheidaei}

   \institute{a) ALBORZ Observatory, Sharif University of Technology,
   Tehran, Iran.\\
   b) Department of physics, Sharif University of Technology,
P.O.Box 11365-9161,Tehran, Iran.}

  \abstract{ A new technique has been devised for the analysis of
extensive air shower data in observing the effect of the moon on
this data. In this technique the number of EAS events with arrival
directions falling in error circles centered about the moving moon
is compared to the mean number of events falling in error circles
with centers randomly chosen in the sky. For any assumed angular
radius of the error circle the deficit in EAS event count in the
direction of moon which is a moon-related effect is interpreted as
the shadow of the moon. A simple theoretical model has been
developed to relate $N_{sky}$ to the angular radius of the error
circle and has been applied to the counts from the moon's
direction in order to extract the physical parameters of the
shadow of the moon. The technique and the
theoretical model has been used on $1.7\times10^{5}$ EAS events recorded at Alborz observatory.\\
   }
   \authorrunning{F. Sheidaei et~al.}
   \titlerunning{On Shadow of the Moon in Extensive Air Shower Data}
 \maketitle
%________________________________________________________________
\section{Introduction}
In 1957, Clark (\cite{clark}) recognized that the moon and the sun
cast a shadow in the isotropic flux of cosmic-ray nuclei. As they
pass overhead during a transit, the moon and the sun block cosmic
rays, so their shadows in the cosmic ray flux should be visible to
Extensive Air Shower (EAS) arrays with sufficiently good angular
resolution. It was not until the early 1990's that an extensive
air shower array had the required angular resolution to observe
this shadow. Since that time the shadow of the sun and the moon
have been used to measure the angular resolution of extensive air
shower arrays (\cite{urban,MACRO,Amenomori}). Observing the shadow
of the moon in EAS experiments which usually might have a much
larger error circle than the disk of the moon is a very difficult
task and requires a careful scrutinization of the data. Normally
study on the shadow of moon is needed as the first step to find
angular resolution and systematic errors of any observational data
which is to be used for any astronomical study in particular for
the research of point sources in sky.
\\
 The object of this paper is to describe a new technique
and procedure devised for the analysis of extensive air shower
data in observing the effect of the moon on this data and also to
present a theoretical model from which we can extract the relevant
physical parameters of the shadow of moon. The technique has been
previously briefly outlined on an investigation of energy spectrum
of EGRET $\gamma-$ray point sources with EAS experiment
(\cite{khakian2}). In Sec.2 we will describe this new technique
and its procedure for EAS data analysis. In Sec.3 we will present
a theoretical model from which we can extract the relevant
physical parameters, In Sec.4 the result of applying the procedure
and the theoretical model to $1.7\times10^5$ EAS data of Alborz
observatory is presented and the obtained results on shadow of the
moon is also presented. Sec.5 is devoted to discussion and some
concluding remarks.
\section{Description of the New Technique and EAS data Analysis procedure}
This technique and the corresponding EAS data analysis procedure
 for observing the effect of the moon on the data and determining
 the pertinent physical parameters is based on corrected measured
 EAS data (corrected for systematic errors). The data must contain
  the following information for each EAS event: The
 arrival time of each shower, $t_{s}$, and the coordinates of the
 arrival direction of the shower. The local
 coordinates of each shower arrival direction should be converted to
  declination and right ascension of each shower with
 arrived direction denoted here by $\delta_{s}$ ,
 $RA_{s}$ respectively. Here, we denote the declination and right
 ascension of the center of the moving moon by $\delta_{m}(t)$ and $RA_{m}(t)$ at the
 arrival time of each shower.
   The EAS experiments generally might have an experimental
   uncertainty which results in the experiment's error circle to
   be  larger than the disk of the moon. Furthermore, normally,
   realistic determination of the radius of the error circle is
   best obtained by observation of the shadow moon in EAS
   data. However, in this new proposed technique and the procedure
   described here, there is no need for the use of a
   predetermined radius of error circle or a fitted value for it,
   and, instead it is based on the observation of any possible
   deficit in shower counts falling in the error circle
   centered about the moving location of moon as compared to
   the average shower counts falling in error circles centered at
   other positions in sky during the observation time for a wide
   range of assumed values for the radius of the error circle
   ranging from $0.2^{\circ}$ to a maximum value relevant to the particular EAS data set
   under analysis. For any assumed radius of circle of error,
   or equivalently, its angular radius $\theta_{err}$ the number of
   showers falling in each circle is determined by
   calculating the angular separations $\theta_{sm}$ between the
   arrival direction of each shower event
   $(\delta_{s},RA_{s})$ and the direction of the center of
   the moon at the time of recording of that event, using the following equation
   from spherical geometry:
\begin{equation}
 \cos\theta_{sm}= \cos \delta_{m} \cos \delta_{s}+
\sin\delta_{m} \sin \delta_{s}\cos(RA_{m}-RA_{s}).
\end{equation}
 Obviously, if $\theta_{sm}<\theta_{err}$ that shower is counted as
falling in the moon's error circle. In order to compare the
obtained result with random sampling and scrutinize the difference
for each assumed value of $\theta_{err}$ some random locations in
the sky denoted by celestial coordinates $(\delta_{r},RA_{r})$ are
chosen and the number of showers falling in the error circles
centered about each of the random locations is determined
similarly by calculating the angular separation $\theta_{sr}$ of
each shower arrival direction $(\delta_{s},RA_{s})$ with the
direction of the center of the randomly chosen error circle
denoted by $(\delta_{r},RA_{r})$ from above equation with
$(\delta_{m},RA_{m})$ replaced by $(\delta_{r},RA_{r})$. If for
any shower event $\theta_{sr}<\theta_{err}$ that shower is counted
as falling in the error circle of that random position.
  For any assumed angular error radius, $\theta_{err}$, some
   error circles are chosen in the sky centered about
   truly random locations. The number of random circles for every angular error radius should be limited such that it ensures
   that no overlap occurs between two or more random circles, so
   the number of random centers are varied from at least 1000 (for small
   error circles) to 77 a smaller number which depends on the data set under analysis (for larger error circles).
   Thus, for each assumed radius for the error circle,
   the mean of the shower counts falling in random circles could safely be
   used as the expected mean number of EAS events falling in the error
   circle in any direction in the moonless sky against which the number of
   EAS falling in the circle centered about the moving moon could
   be safely compared, and, the variance of its distribution could
   safely be used as the statistical error of the mean number.
   Obviously if the deficit in the number of showers falling in
   each error circle centered about the moving moon from the mean
   number exceeds the statistical uncertainly in the mean
   , then the deficit could only be attributed to the moon's effect. This effect is the shadow of
   moon in the EAS data, and, as explained in Sec.3, from a
   quantitative analysis and comparison with the expected mean number
   of EAS showers falling in the randomly centered error circle
   with that falling in the circles centered about the moving
   moon, the physical parameters of the moon's shadow could be
   extracted.
\section{A theoretical model for Moon's shadow in EAS data}
 Following the procedure described in above technique, one now has
the expected mean number of showers falling in randomly chosen
error circles in monless sky as well as that in the moon-centered
error circle for a set of assumed radii of the error circles,
$\rho_{err}$, (or equivalently angular radii, $\theta_{err}$).
Here we propose a simple physical model to obtain this expected
mean number as a function of the angular radius of the error
circle, as a function of $\theta_{err}$. We now derive the
expected mean number of EAS events from each random direction in
the moonless sky as a function of the assumed radius for the error
circle, $N_{sky}(\rho_{err})$. The derivation is based on the
single assumption of the model, that is, the assumption of uniform
intensity, I, of EAS producing radiation everywhere in the $4\pi$
steradian for each area of the random moonless sky and thus for
each element of the error circle. For the contribution from each
element of area of the error circle ($2\pi\rho d\rho$), we should
take into account only a fraction of radiations coming at such an
angle to be able to reach the point of observation, that is a
fraction equal to $\frac{Id\Omega}{4\pi}$. Where $d\Omega$ is the
solid angle subtended by the element of area element from the
observation point, which is the projection of the area element
divided by the square of its distance from observation point, that
is, $d\Omega=2\pi\rho d\rho(\frac{d}{R})\frac{1}{R^{2}}$, with
$R=\sqrt{\rho^{2}+d^{2}}$ and d is the distance from point of
observation to the center of the error circle, and it is merely a
multiplicative constant factor relating the radius of the error
circle to its angular radius $\rho_{err}=d \tan \theta_{err}$. The
integration is trivial. Thus we have:
\begin{equation}
N_{sky}(\rho_{err})=\int^{\rho_{err}}_{0}\frac{Id}{4\pi}\frac{2\pi\rho
d \rho}{[\rho^{2}+d^{2}]^{\frac{3}{2}}}
=-\frac{Id}{2}[\rho^{2}+d^{2}]^{-\frac{1}{2}}\mid^{\rho_{err}}_{0}\\
=-\frac{Id}{2}(\frac{1}{\sqrt{\rho_{err}^{2}+d^2}}-\frac{1}{d})
\end{equation}
 for the number of showers falling in the error circle
centered about moon, $N_{moon}(\rho_{err})$ the integration has to
be split into two or tree parts involving the physical parameters
of shadow of moon in EAS data. Here, we define the following three
physical parameters used in this model:
 \\
 a) $r_m\equiv$ radius of umbra  of shadow that is, from
 $\rho=0$ to $\rho=r_m$ the EAS producing radiation are assumed to be fully
 absorbed (totally blocked) and could not contribute to $N_{moon}(\rho_{err})$.
 \\
 b) $r_{p}\equiv$ radius of penumbra of shadow; that is from $\rho=r_m$
 to $r=r_p$ only a fraction (\emph{f}) of EAS producing radiation
 penetrate the penumbra and contribute to $N_{moon}(\rho_{err})$.
\\
c) $f\equiv$ the fraction of EAS producing radiation which
 penetrate the moon's penumbra.
 \\
Obviously, if $\rho_{err}\leq r_p$ the integration will only be
 split into two parts, that is, $ \int^{\rho_{err}}_{0}\rightarrow 0\times\int^{r_{m}}_{0}+ f \times
 \int^{\rho_{err}}_{r_{m}}$.
 \\
 For the case of $\rho_{err}>r_p$, the integral will be split
 into three parts:
 \\
 $\int^{\rho_{err}}_{0}\rightarrow 0\times\int^{r_{m}}_{0}+f \times
\int_{r_{m}}^{r_{p}}+ 1\times\int^{\rho_{err}}_{r_{p}}.$
\\
The result of these integrations giving $N_{moon}(\rho_{err})$ in
terms of physical parameters of the moon's shadow $(r_m,r_p, f)$
is given in Appendix. It should be emphasized here that the strict
explicit assumption of uniform flux of EAS producing radiation
used in this model requires that when the EAS data is used to
extract the parameters of the moon's shadow from this model one
has to make sure that the data may only have statistical error (as
this has also been implicitly assumed as outlined in the procedure
for obtaining shower counts in the error circles), that is, the
data should have been corrected for any systematic errors such as
those related to the site of observation and non- Uniformities in
the exposure time in various directions in sky.
\section{Application of the Technique to ALBORZ EAS data}
The technique described in Sec.2 for determination of the moon
shadow has been applied to $1.7\times10^5$ EAS data collected in
280 hours of observations in April-June 2002, with the small EAS
array of the prototype of ALBORZ Observatory of Sharif University
located in Tehran, Iran (51$^{\circ}$~20$^{'}$E and
35$^{\circ}$~43$^{'}$N, elevation 1200~m
$\equiv$~890~g~cm$^{-2}$). For details of array and data, see
(\cite{Bahmanabadi2}). As explained in our previous
report(\cite{Khakian}), the data has been corrected (scaled for
uniform exposure) for site-dependent factors effecting shower
counts from different directions in sky. The information on the
celestial coordinates of the moon during the observation time of
the collected data has been obtained from the internet site
(http://aa.usno.navy.mil). The moon's data has been obtained for
time increments of one minute, and the location of moon in Right
ascension and declination coordinate at the recorded time of
arrival of each EAS event has been calculated and used.\\
 For an assumed set of radii for error circles ranging from $0^\circ$ to
$6^\circ$ (increments of $0.2^\circ$) we have calculated the
number of corrected shower counts falling in each error circle.
For every assumed radius, some random moon-like locations passing
through the paths like as moon's path in sky was chosen according
to the radius of error circles and the number of corrected EAS
counts falling in each error circle was found according to the
procedure described in Sec.2. In table 1 the number of random
circles and the mean of counts for various assumed radii of error
circles are given in second and third columns, also in this table
(4th column) the number of corrected shower counts from the error
circle centered about the moving moon is shown. In column 5 and 6
of the table the deficit of counts from moon's error circle from
the mean count of the random sky error circle and the statistical
error of deficits are given. The last column of the table gives
the statistical significance of these deficits calculated with
Li\&Ma method (\cite{lima}). Fig.1 shows the variation of the mean
number of events for moon-like error circles with random centers
as a function of the chosen radius of the error circles. The
smooth curve shown that calculated according to our theoretical
model of Sec.3 and it fits the mean count from random sky with a
regression of 0.996. Error bars are taken from 4th column of table
1. The good fit of random sky counts with model shows that we can
safely use these mean number of events to compare with that
falling in the error circles centered about the moving moon and
rule out the possibility of the deficit in the number of events
falling in the moon centered circle as due to statistical
fluctuations.In Fig.2 we have shown the variation of events
falling in each circle centered about the moon and the mean number
of events falling in the error circles centered about random
moving moon-like locations as a function of radius of the error
circles. As seen in Fig. 2 moon counts are less than mean counts
from random moon-like centers for all error circles that we
considered. In Fig.3 we have shown the number of deficit events
for each radius of the error circles.
 We have fitted the deficit counts falling in the error
circle centered about the moving moon from that for moon-like
circles with random centers to our theoretical model (sec.3 and Eq. A2 in Appendix)
and have obtained the following results: \\
$\theta_{m}=0.5^\circ,\theta_{p}=4.5^\circ,f=0.80$.
\section{Concluding Remarks}
It is worth remarking that the application of the proposed
technique to ALBORZ EAS data has yielded good agreement between
the mean number of counts from error circles with centers chosen
randomly in sky with no moon in the line of sight and the expected
number according to our proposed theoretical model. This good
agreement is very encouraging and prompted us to extract the
physical parameters of the moon shadow (defined in Sec.3) from
this data. It should also be remarked that the data used for
calculating shower counts in each error circle was the corrected
counts scaled in order to obtain a uniform exposure of sky. The
correction accounted for site-dependent systematic errors arising
from uneven number of EAS events in various directions in sky due
to two main factors: (1) varying amount of air mass which produces
the EAS event as a function of zenith angle and depends on the
elevation of the site (\cite{Khakian}), and (2) geomagnetic effect
which depends on the components of magnetic field at the site's
location. Our attempt to extract the physical parameters of moon's
shadow from this data has been fully successful as can be seen
from the reported result in Sec.4. That is in fitting the
corrected data to our theoretical model we are able to obtain a
value for the radius of shadow's umbra $\theta_m=0.5^{\circ}$.
However , according to statistical significance shown in the last
column of table 1 we didn't see the umbra with good significance
but the obtained results show that in spite of low-statistics EAS
data base this method is powerful to find shadow of moon. One may
suggest that the value of $\theta_p=4.5^\circ$ we obtained is just
the umbra's radius rather than penumbra's and resulted from low
angular resolution of our array . This could be right since the
angular resolution of our array which was reported before
(\cite{khakian2}) is about $4.3^{\circ}$ close to $4.5^{\circ}$
which we find here as the radius of penumbra. We believe that the
main uncertainty in extracting results from Alborz data could be
due to the following two reasons, both of which
will be improved upon in future with much higher number of events and larger statistics. \\
\\
\textbf{1. Inaccurate and low-statistics EAS data base}\\
 Since the umbra's radius is in the order of $0.5^\circ$ it is hard to expect to extract it from inaccurate EAS
data. The local coordinates associated with each EAS event in
ALBORZ EAS data has been obtained from an array with a very small
number of detectors. EAS data from observatories with large
arrays, once corrected for the systematic site-dependent errors
may be more accurate to yield better results for the physical
parameters of  moon's shadow according
to the technique presented here.\\
\textbf{2. Incomplete data on Moon}\\
As explained in Sec.3 the information about the celestial
coordinates of the moon was obtained from the internet site using
time increments of one minute. The time of arrival of EAS events
had been recorded with an uncertainty of 0.07 seconds. In the
computations of shower counts in the error circles of various
radii centered about the moving moon which are given in Table 1 to
check whether a given EAS event falls in the error circle centered
about the moon or falls outside it, we have used the coordinates
of the moon at the one of the minute steps which is closest to the
arrival time of the given EAS event. Obviously, this may have
caused an extra inaccuracy in the counts given in column 5 of
Table 1. In future application of this technique the interpolated
or exact location of the moon at the instant of recording of each
EAS event must be used and the variable earth-moon distance should
also be taken into account. However, the study of the moon's
motion has been beyond the scope of the present work.
\section{Appendix}
For calculating the count in the error circle centered about the
moon, we split Eqn.2 in three parts. The result of integrations
for two regions are shown in following equations: \\
$N_{moon}=-\frac{I}{2}[(\cos\theta_{err}-cos\theta_{p})+f(\cos(\theta_{p}-\cos\theta_{m}))],\hspace{0.5cm}\theta_{err}>\theta_{p}\hspace{2cm}
$\\
$N_{moon}=-\frac{I}{2}.f(\cos\theta_{err}-cos\theta_{m}),\hspace{2.7cm}\theta_m<\theta_{err}<\theta_{p}\hspace{2.0cm}
(A.1)$\\
If we now subtract the number of events in the random circles,
$N_{sky}$(Eq.2) from above we obtain:\\
$N_{back}-N_{moon}=-\frac{I}{2}[\cos\theta_{err}-1] \hspace{3.8cm}\theta_{err}<\theta_{m}$\\
$N_{back}-N_{moon}=-\frac{I}{2}[(1-f)\cos\theta_{err}-1+fcos\theta_m]
\hspace{1cm}\theta_m<\theta_{err}<\theta_{p}\hspace{1cm}(A.2)
$\\
$N_{back}-N_{moon}=-\frac{I}{2}[(1-f)\cos\theta_p-1+f\cos\theta_m]\hspace{1cm}\theta_{p}<\theta_{err}
$\\
The parameter \emph{I}(=497270) is determined by fitting Eqn.2 to
the data of column 3 in table 1. By knowing \emph{I} and fitting
the data of deficit events (column 6 of table 1) with above
equations (A.2), we obtained $\theta_m,\theta_p,$ and \emph{f}.
\section*{Acknowledgement}
\small{This research was supported by a grant No. NRCI 1853 from
the national research council of Islamic republic of Iran for
basic sciences.}
%%Format tables as in the following example

\begin{table}
\tiny{
\begin{tabular}{|c|c|c|c|c|c|}
  \hline
  \hspace*{4.0mm}radii of & \hspace{0.2cm}&Circle centered&Deficit of counts& \hspace{0.2cm}error in&\hspace{0.2cm}\,\hspace{0.1cm}Statistical\hspace{0.2cm}\, \\
  error circle$(^\circ)$&\hspace{0.4cm}\tiny{random moon-like locations}\hspace{0.3cm}&about the moon\hspace{3cm}&from the moon&deficit&\hspace{0.2cm}Significant\hspace{0.3cm}\\
  \hline
  \end{tabular}
  \\
  \begin{tabular}{|c|c|c|c|c|c|c|}
\hline
 \hspace{1.5cm}\,&\tiny{\#random circles}&\tiny{mean count}&\hspace{0.6cm}\tiny{counts}\hspace{.4cm}\,&\hspace{0.3cm}\,\tiny{counts}\hspace{.8cm}\,&\tiny{in counts}&\tiny{Li\&Ma Method}\\
 \hline
 0.1&1000&7&5&2&2.24&0.89\\
0.3&1000&15&6&9&2.46&3.67\\
0.5&1000&26&17&9&4.13&2.18\\
0.7&1000&40&22&18&4.70&3.83\\
0.9&1000&55&42&13&6.50&2.0\\
1.1&1000&74&52&22&7.23&3.05\\
1.3&1000&97&73&24&8.56&2.81\\
1.5&1000&125&93&32&9.67&3.31\\
1.7&991&155&105&50&10.29&4.87\\
1.9&793&187&134&53&11.63&4.57\\
2.1&649&223&168&55&13.03&4.24\\
2.3&541&261&204&57&14.38&3.98\\
2.5&458&303&246&57&15.82&3.63\\
2.7&393&348&287&61&17.09&3.59\\
2.9&340&396&317&79&17.99&4.43\\
3.1&298&446&365&81&19.36&4.23\\
3.3&263&497&422&75&20.81&3.64\\
3.5&233&553&454&99&21.67&4.63\\
3.7&209&610&487&123&22.56&5.56\\
3.9&188&669&518&151&23.39&6.61\\
4.1&170&727&579&148&24.77&6.13\\
4.3&155&791&636&155&26.03&6.12\\
4.5&141&858&709&149&27.52&5.57\\
4.7&129&924&780&144&29.03&5.13\\
4.9&119&994&840&154&30.31&5.29\\
5.1&110&1068&924&144&31.88&4.71\\
5.3&102&1140&972&168&32.84&5.36\\
5.5&94&1210&1048&162&34.32&4.97\\
5.7&88&1286&1132&154&35.93&4.55\\
5.9&82&1364&1205&159&37.31&4.55\\
6.1&77&1434&1272&162&38.78&4.51\\
 \hline
 \end{tabular}}
 \small{\caption{Number of EAS events obtained in various error circles with random centers and moon center,
 of the low-statistics EAS data of Alborz observatory.}}
\end{table}
\begin{figure}[!htb]
\begin{center}
\includegraphics[width=100mm,height=60mm]{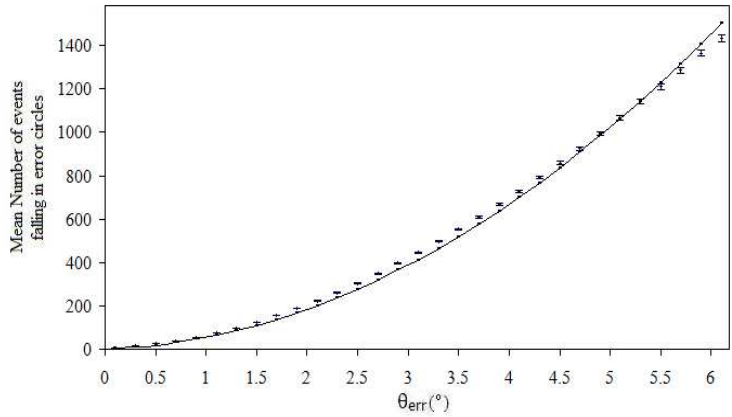}%
\end{center}
 \caption{Variation of the mean number of EAS events falling in the random error
 circles as a function of the angular radius of the error circle, $\theta_{err}$. The smooth curve is the result of
computations according to our theoretical model (Sec.2).}
% \label{fit}
 \end{figure}
\begin{figure}
   \centering
   \includegraphics[height=7cm,width=10cm]{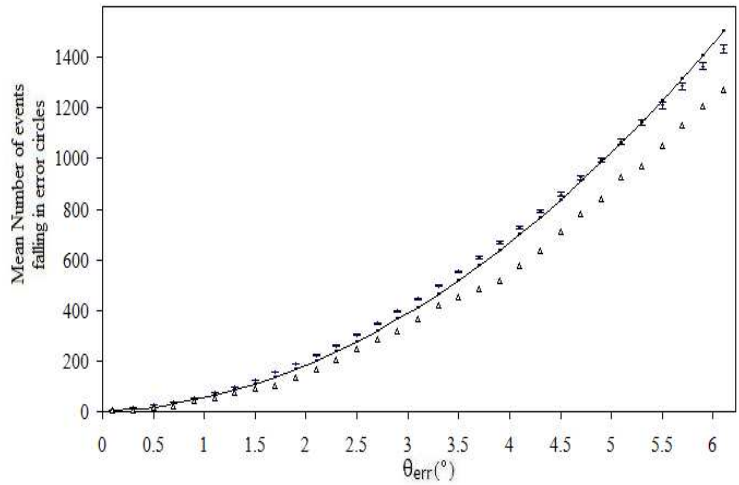}
      \caption{Variation of events falling in the random circles($\bullet$) and moving moon($\bigtriangleup$) as a
      function of the angular radius of the error circle, the smooth curve is the same as Fig.1.}
         \label{setup}
   \end{figure}
\begin{figure}
   \centering
   \includegraphics[height=5cm,width=10cm]{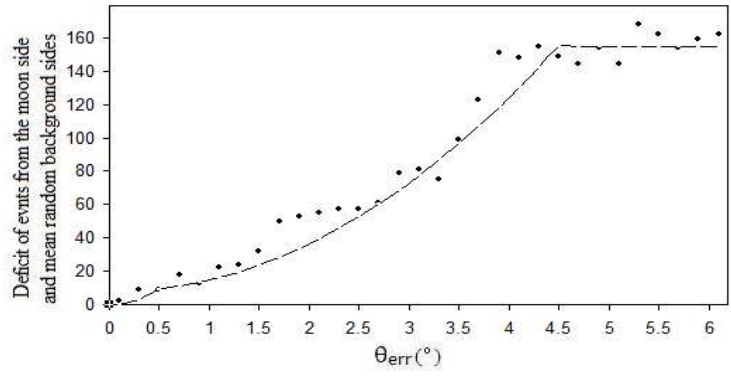}
      \caption{Variation of deficit events falling in moving moon circles from that in  the random circles as a function of the angular radius of the error circle, the smooth two parts curve is
      the result of fitting data with the theoretical model equations see in appendix.}
         \label{setup}
   \end{figure}
\end{document}